%% file: arxiv.tex
\pdfoutput=1
\documentclass[nonacm]{acmart}
\setcopyright{none}
\acmDOI{}
\acmPrice{}
\acmISBN{}
\usepackage[utf8]{inputenc}

\usepackage{xspace}
\usepackage{multirow}
\usepackage{graphicx}
\usepackage{todonotes}
\usepackage{subfigure}
\usepackage{listings}
\usepackage{url}
\usepackage{balance}
\usepackage{paralist}
\usepackage[ruled,vlined,linesnumbered]{algorithm2e}
\newcommand{\outert}{\textit{outer}\xspace}
\newcommand{\inner}{\textit{inner}\xspace}

\newcommand{\ARM}{\textsc{CTE-ARM}\xspace}

\newcommand{\NESTED}{\textsc{Nested}\xspace}
\newcommand{\NONESTED}{\textsc{Non-nested}\xspace}
\newcommand{\OPT}{\textsc{Opt-D}\xspace}
\newcommand{\OPTIF}{\textsc{Opt-D-Cost}\xspace}

\newcommand{\MTBLAS}{\mbox{mt-BLAS}\xspace}

\newcommand{\boneS}{\texttt{boneS10}\xspace}
\newcommand{\inline}{\texttt{inline\_1}\xspace}
\newcommand{\bone}{\texttt{bone010}\xspace}

\newcommand{\audi}{\texttt{audikw\_1}\xspace}
\newcommand{\Gcircuit}{\texttt{G3\_circuit}\xspace}
\newcommand{\thermal}{\texttt{thermal2}\xspace}

\newcommand{\sdkq}{\texttt{s3dkq4m2}\xspace}

\title{Optimization of the Sparse Multi-Threaded Cholesky Factorization for A64FX} 

\author{Valentin Le Fèvre}
\affiliation{\institution{Barcelona Supercomputing Center}
\city{Barcelona}
\country{Spain}}
\email{valentin.lefevre@bsc.es}
\author{Tetsuzo Usui}
\affiliation{\department{Next Generation Technical Computing Unit}
\institution{Fujitsu Limited}
\city{Kawasaki}
\country{Japan}}
\email{t-usui@fujitsu.com}
\author{Marc Casas}
\affiliation{\institution{Barcelona Supercomputing Center}
\city{Barcelona}
\country{Spain}}
\email{marc.casas@bsc.es}

\graphicspath{{figs/}}

\begin{abstract}
Sparse linear algebra routines are fundamental building blocks of a large variety of scientific applications. 
Direct solvers, which are methods for solving linear systems via the factorization of matrices into products of triangular matrices, are commonly used in many contexts.
The Cholesky factorization is the fastest direct method for symmetric and definite positive matrices.
This paper presents selective nesting, a method to determine
the optimal task granularity for the parallel Cholesky factorization based on the structure of
sparse matrices. We propose the OPT-D-COST algorithm,
which automatically and dynamically applies selective nesting.
OPT-D-COST 
leverages matrix sparsity to drive complex task-
based parallel workloads in the context of direct solvers.
We run an extensive evaluation campaign considering a heterogeneous set of 60 sparse matrices and a parallel machine featuring the A64FX processor.
OPT-D-COST delivers an average performance speedup of 1.46$\times$ with respect to the best state-of-the-art parallel method to run direct solvers.
\end{abstract}
\keywords{cholesky, sparse matrix, multi-threading, parallel tasks, granularity}

\begin{document}

\maketitle


\section{Introduction}
\label{sec.intro}

\input{intro.tex}
%

\section{Background on CHOLMOD} 
\label{sec.cholmod}

\input{cholmod.tex}

\section{Task-Based Parallel CHOLMOD}
\label{sec.parallel}

\input{parallel.tex}

\section{Sparsity-Driven Selective Nesting}
\label{sec.nesting}

\input{nesting.tex}

\section{Evaluation}
\label{sec.expe}

\input{expe_arxiv.tex}

\section{Related work}
\label{sec.related}
\input{related.tex}

\section{Conclusion}
\label{sec.conclusion}

\input{conclusion.tex}

\balance
\bibliography{biblio.bib}
\bibliographystyle{plain}

\end{document}

%% file: intro.tex
Linear algebra routines are common building blocks of scientific computing workloads. 
In particular, direct linear solvers are widely used.
Indeed, a very common way to solve a system such as $Ax=b$
is to factorize matrix $A$ into a product of triangular matrices.
When dealing with symmetric and definite positive matrices, the Cholesky factorization~(\cite{hornJohnsonMatrix}, Corollary 7.2.9) is faster than alternative direct methods like the LU~(\cite{hornJohnsonMatrix}, Definition 3.5.1) or the QR~(\cite{hornJohnsonMatrix}, Theorem 2.1.14) factorizations.
The Cholesky decomposition factorizes matrix $A$ in terms of a product of a lower triangular matrix $L$ and its transposed matrix $L^T$,  that is, $A = LL^T$.

For the case of sparse matrices, there are several high-performance parallel solvers implementing the Cholesky factorization: MUMPS~\cite{mumps}, \textsc{PaStiX}~\cite{pastix}, PARDISO~\cite{PARDISO} or CHOLMOD~\cite{cholmod}.
CHOLMOD is one of the most widely used implementations of the Cholesky factorization.
It is an open source software.
In particular, CHOLMOD implements a supernodal version of the Cholesky factorization.
Supernodal algorithms rely on
a decomposition of the matrix in terms of blocks of columns, that we call supernodes.
These algorithms rely on sub-operations (Level-3 BLAS kernels) on a supernode or between supernodes to perform the factorization of the whole matrix.
Like many numerical solvers, CHOLMOD is composed of three main computing phases: The first one \textit{analyzes} the structure of the matrix and reorders matrix rows and columns to minimize the fill-in, i.e. the number of non-zero elements in the final factorization.
The second phase
\textit{factorizes} the matrix by applying BLAS kernels within a supernode or between them, and the third one \textit{solves} a couple of triangular systems of equations, namely
$Ly = b$ and $L^T x = y$ in the case of Cholesky factorization. 
Among the three phases, the \textit{factorize} phase represents the bulk of the CHOLMOD solver.

Although the state-of-the-art CHOLMOD implementation is sequential, one can easily link it with multi-threaded BLAS libraries~\cite{BLAS} to take advantage of many-core architectures.
Other approaches rely on task-level parallelism to run direct solvers on sparse matrices~\cite{tangHybrid,pastixRuntime}.
They exploit the structure of the elimination tree~\cite{liuTree}, which represents the order in which supernodes have to be factorized, to orchestrate the parallel run.
Other aspects of the matrix sparse pattern are usually ignored by previous works.
Since the computational load of every task heavily depends on the structure of its corresponding supernode, current schemes are prone to load balancing problems.
More advanced approaches~\cite{mumpsQR}, consider different features of the input matrix to optimize the setup of the parallel run in terms of aspects like task granularity, but determining the quality of these parameter setups requires trying them all.

This paper presents \textit{selective nesting}, a method to determine the optimal task granularity based on the structure of sparse matrices.
We propose the \OPTIF algorithm, which automatically and dynamically applies selective nesting. 
The main criteria
that \OPTIF uses to determine task granularity is the number of inter-node operations per intra-node operation.
Applying \OPTIF on a heterogeneous set of 60 sparse matrices yields an average speed-up with respect to state-of-the-art multi-threaded BLAS CHOLMOD 
of 1.95$\times$.
When comparing \OPTIF with state-of-the-art task-based approaches for direct solvers, it obtains average speed-ups of 1.46$\times$.
\OPTIF is the first approach to leverage matrix sparsity to drive complex task-based parallel workloads in the context of direct solvers. 
This information increases load balancing, reduces task creation overhead, and significantly improves performance with respect to state-of-the-art approaches.

The remainder of the paper is organized as follows:
Section~\ref{sec.cholmod} summarizes the sequential supernodal CHOLMOD algorithm.
Section~\ref{sec.parallel} presents our baseline task-based parallelization of CHOLMOD using OpenMP. 
Section~\ref{sec.nesting} analyzes several heuristics to determine task granularity.
We evaluate the different algorithms on a wide and heterogeneous set of matrices using A64FX processors
in Section~\ref{sec.expe}. 
Section~\ref{sec.related} describes relevant state-of-the-art approaches and, finally, we provide some
concluding remarks in Section~\ref{sec.conclusion}.

%% file: cholmod.tex
This section briefly describes the supernodal CHOLMOD algorithm~\cite{cholmod}. 
The default CHOLMOD algorithm is composed of three main phases: analysis, factorization, solve.
The analysis phase traverses the input matrix and applies optimizations like matrix reordering, which minimizes the number of non-zeros in the final factorization,
elimination tree setup~\cite{liuTree}, which determines data-dependencies between partial factorizations, 
and node amalgamation, which merges nodes of the elimination tree corresponding to adjacent columns and obtains what we call \textit{supernodes}.
Matrix reordering is an important performance factor, as with less non-zeros, both the number of operations and the memory consumption are reduced. In CHOLMOD, several algorithms like METIS~\cite{metis} or AMD~\cite{amd} are used
and the best ordering method is kept.
These algorithms are heuristics, since matrix reordering is a NP-complete problem~\cite{fillin_np}.
Node amalgamation
makes Level-3 BLAS and LAPACK routines more efficient when applied to supernodes.
The denser and larger the supernode is, the more efficient BLAS routines become in terms of cache locality and data reuse.
Large nodes may incur some overhead in terms of redundant operations involving zero values, which wastes execution time and memory.

Due to the short duration of the initial phase, this paper is not focused on its improvement, although some of its optimizations impact the algorithms we discuss in Section~\ref{sec.nesting}.
For example, node amalgamation determines supernodes count and size. 
The size of the inter-node operations grows with the supernode size, and the number of calls to BLAS kernels increases with the supernode count.
The final solve phase obtains the final solution of the linear system by solving a couple of triangular systems. 
This paper does not apply any optimization to the solve phase since it is short and simple.

%% file: parallel.tex
This section describes our parallelization of the factorization phase of CHOLMOD, which is based on task parallelism.
This baseline implementation is equivalent to state-of-the-art parallel approaches for direct solvers~\cite{tangHybrid, pastixRuntime}.
We use the task constructs of the OpenMP programming model.

The initial structure of the factorization algorithm consists in a for loop over all supernodes in ascending order (i.e. starting with the supernode who has the leftmost columns and finishing with the rightmost supernode).
We refer to this loop as the main loop.
For each supernode, the updates coming from previous supernodes are processed using the SYRK and GEMM BLAS kernels.
Then, the supernode is factorized using LAPACK's POTRF and TRSM from BLAS. This code is said to be \textit{left-looking}, meaning that each supernode accumulates updates from previous (i.e. on its left) supernodes
before factorizing. 
Note that this approach is very close to tiled algorithms for dense matrices~\cite{cholesky-tile}, except that all sub-diagonal tiles are merged together in a supernode.
While the factorization of the supernode is composed of only two kernel calls (POTRF and TRSM), the updates can come from any of the previous supernodes. 
This is translated as
another for loop inside the main loop, that we refer as the inner loop.

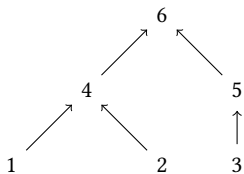
\begin{figure}
\begin{center}
 \begin{tikzpicture}
  \node[circle] (6) at (6,6) {6};
  \node[circle] (5) at (7,5) {5};
  \node[circle] (4) at (5,5) {4};
  \node[circle] (3) at (7,4) {3};
  \node[circle] (1) at (4,4) {1};
  \node[circle] (2) at (6,4) {2};
  \draw[->] (1) -- (4);
  \draw[->] (2) -- (4);
  \draw[->] (3) -- (5);
  \draw[->] (5) -- (6);
  \draw[->] (4) -- (6);
 \end{tikzpicture}
 \end{center}
 \caption{A simple elimination tree with 6 supernodes.\label{fig.tree}}
\end{figure}

Our strategy to parallelize CHOLMOD consists in exploiting the information available in the elimination tree to orchestrate the parallel run.
While supernodes are factorized one after another in the baseline CHOLMOD code, it is possible to factorize in parallel some supernodes. 
We illustrate this parallel factorization with a simple example.
Figure~\ref{fig.tree} shows a simple elimination tree, where each node label corresponds to the index of the supernode. 
CHOLMOD
first factorizes supernode 1, then 2, and so on until factorizing supernode 6. 
However, this tree clearly shows that supernode 1 and 2 can be factorized in parallel. 
Supernode 4 needs both supernodes 1 and 2 to be factorized before it can be processed.
Similarly, supernodes 3 and 5 can be factorized in parallel of supernodes 1, 2 and 4. 
An optimal parallel algorithm should factorize supernodes 1, 2 and 3 in parallel, then 4 and 5 in parallel, and finally 6. 
Our strategy is based on a task-based parallel execution model where tasks represent the factorization of
one supernode, and dependencies between them are directly given by the elimination tree.
It is also possible to use multi-threaded versions of BLAS and LAPACK kernels to run CHOLMOD in parallel.
We consider and evaluate both strategies in Section~\ref{sec.expe}.


Using OpenMP \texttt{\#pragma omp task} directives, we can specify each iteration of the main loop as one factorization task. We define these tasks as \textit{\outert tasks}.
Each one of these tasks execute all BLAS and LAPACK kernels related to the factorization of its supernode.
Most elimination trees can be decomposed in linear chains that join at some point. 
At the beginning of the parallel run, they can expose a large amount of parallelism because many supernodes can be independently factorized.
However, near the end of the matrix factorization, there is a reduced number of independent supernodes.
Previous work discusses trade-offs between elimination tree parallelism and BLAS parallelism for the QR factorization~\cite{buttariQR}.

To leverage as much concurrency as possible, we define nested tasks inside the source code of the \textit{\outert tasks}.
We call them \textit{\inner tasks}.
The easiest way of parallelizing one \outert task is to separate the updates (SYRK+GEMM) that occur from previous supernodes. 
We add another \texttt{\#pragma omp task} directive that will
encapsulate the content of the inner loop. 
These updates are not totally parallel since, despite the fact that the computation itself is independent
for each previous supernode, partial outputs need to be assembled within the \outert task at the end. 
These write operations must not take place simultaneously to enforce correctness.
We make use of OpenMP locks to prevent concurrent writing in this part. 
Alternative approaches to avoid simultaneous writes to the same address, like using the OpenMP task depend
clause, incur significant overhead for these fine-grain updates and force the programmer to statically specify an update order, which might undermine the performance.

Listing~\ref{algo.main} summarizes the parallel factorization algorithm. 
Here, \texttt{dep\_in} and \texttt{dep\_out} are arrays that indicate the input and output dependencies
for all supernodes. 
There are no dependencies across the \inner tasks, just a serialized write operation at the end.
Each \outert task contains a final synchronization point in terms of a \texttt{\#pragma omp taskwait}  
to enforce the completion of all \inner tasks before finalizing the \outert task. 
Indeed, all updates must be completed before we can factorize the node. 
This code constitutes the baseline task-based CHOLMOD. Section~\ref{sec.nesting} describes sparsity-driven approaches to drive parallel granularity.

%

\begin{minipage}{0.9\linewidth}
\begin{lstlisting}[caption={Structure of the implementation of \outert and \inner tasks in CHOLMOD with OpenMP directives.},numbers=left,numbersep=0.5em,label={algo.main},captionpos=b,basicstyle=\tiny,escapechar=|]
 for (s = 0 ; s < nsuper ; s++)
 {
    #pragma omp task in({*(dep_in[s][ii]), ii=0;num_in[s]}) out(*dep_out[s]) \
    default(none) shared(...) firstprivate(...) private(...) label(outer)
    {
       //Construction of the supernode
       for (idxS = STp [s] ; idxS < STp[s+1] ; idxS++) { |\label{line.inner_loop}|
            d = STi[idxS] ;
            if (d==s) continue ;
	    #pragma omp task default(none) shared(...) \
	    private(...) firstprivate(...) private(...) label(inner)
	    {
		//SYRK and GEMM
		omp_set_lock(&omp_lock);
		//Assembly of supernode
		omp_unset_lock(&omp_lock);
	    }
       }
       #pragma omp taskwait
       //POTRF and TRSM
     }
 }
\end{lstlisting}
\end{minipage}

%% file: nesting.tex
This section presents the selective nesting concept, a method to determine the optimal task granularity based on the matrix sparsity.
We present two heuristics to drive the use of nested tasks for the parallel Cholesky factorization. 
Subsection~\ref{sec.nesting.simple} motivates, through some simple examples, the use of selective nesting.
Subsection~\ref{sec.nesting.opt} describes an auto-adaptive efficient algorithm, derived from the prior analysis. 
Subsection~\ref{sec.nesting.optif} extends this auto-adaptive algorithm by considering the cost
of \inner tasks.
Finally, we consider a hybrid approach based on a simple condition to get the best of both our heuristics and 
the existing multi-threaded CHOLMOD algorithm in Subsection~\ref{sec.nesting.blas}.

\subsection{Why is selective nesting necessary?}
\label{sec.nesting.simple}

\begin{figure}
\subfigure[Without nested tasks (\NONESTED)]{ \includegraphics[width=\linewidth,height=1.2cm]{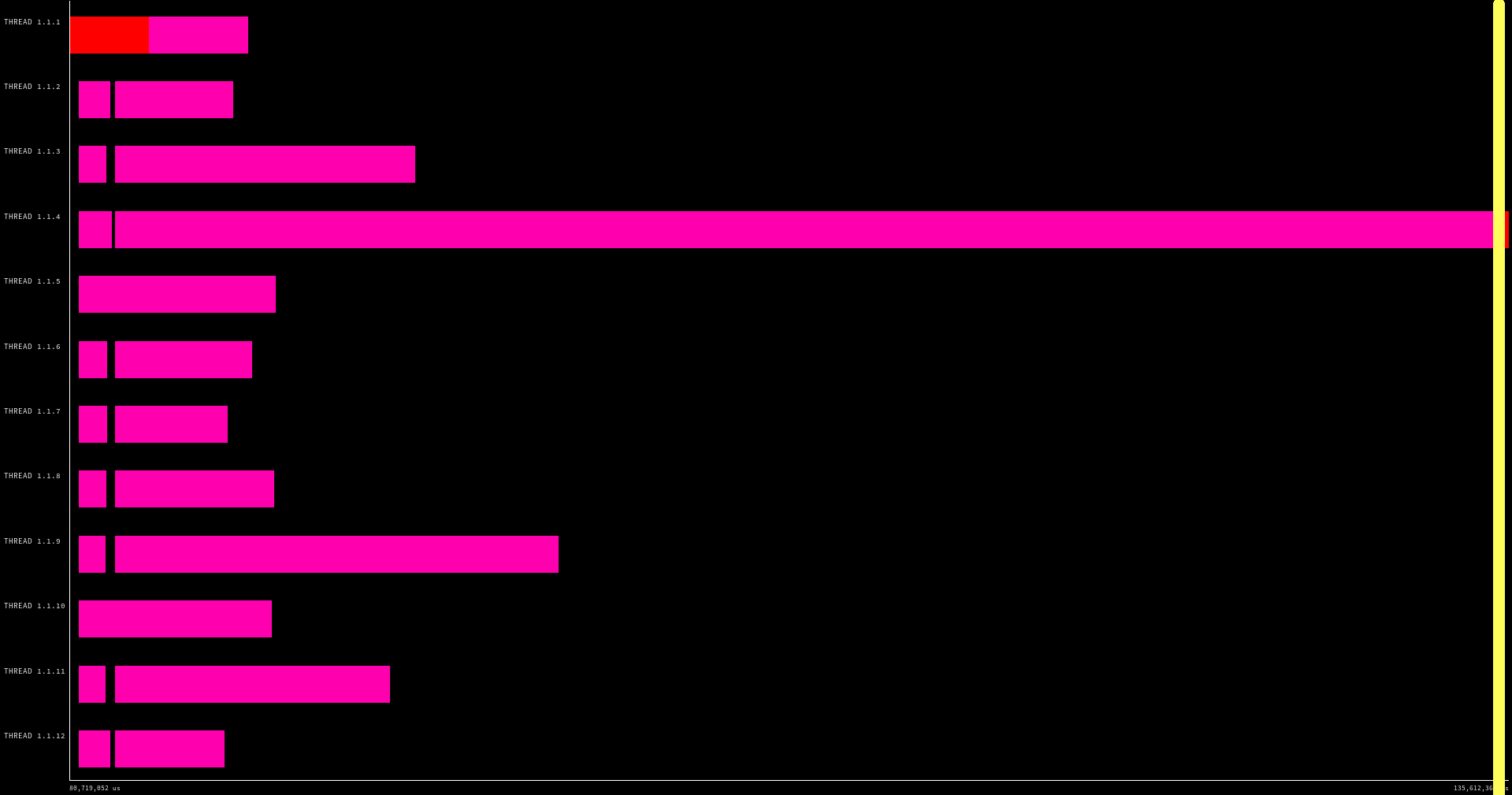}}
 \subfigure[With nested tasks (\NESTED)]{ \includegraphics[width=\linewidth,height=1.2cm]{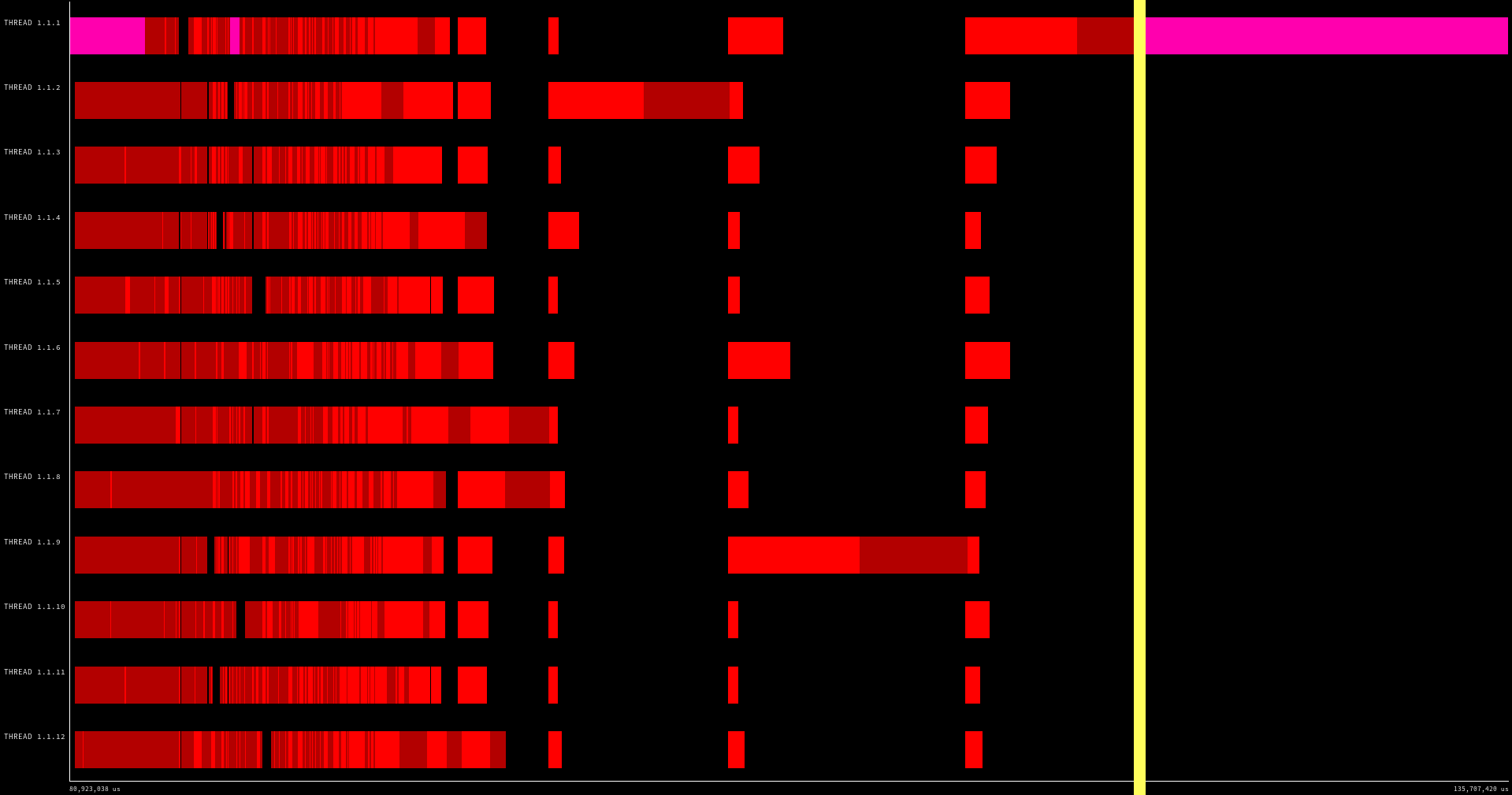}}
 \caption{Comparison of execution traces with or without nested tasks for \bone. The yellow line indicates the end of the parallel part.\label{fig.nesting_bone010}}
\end{figure}

\begin{figure}
 \subfigure[Without nested tasks (\NONESTED)]{ \includegraphics[width=\linewidth,height=1.2cm]{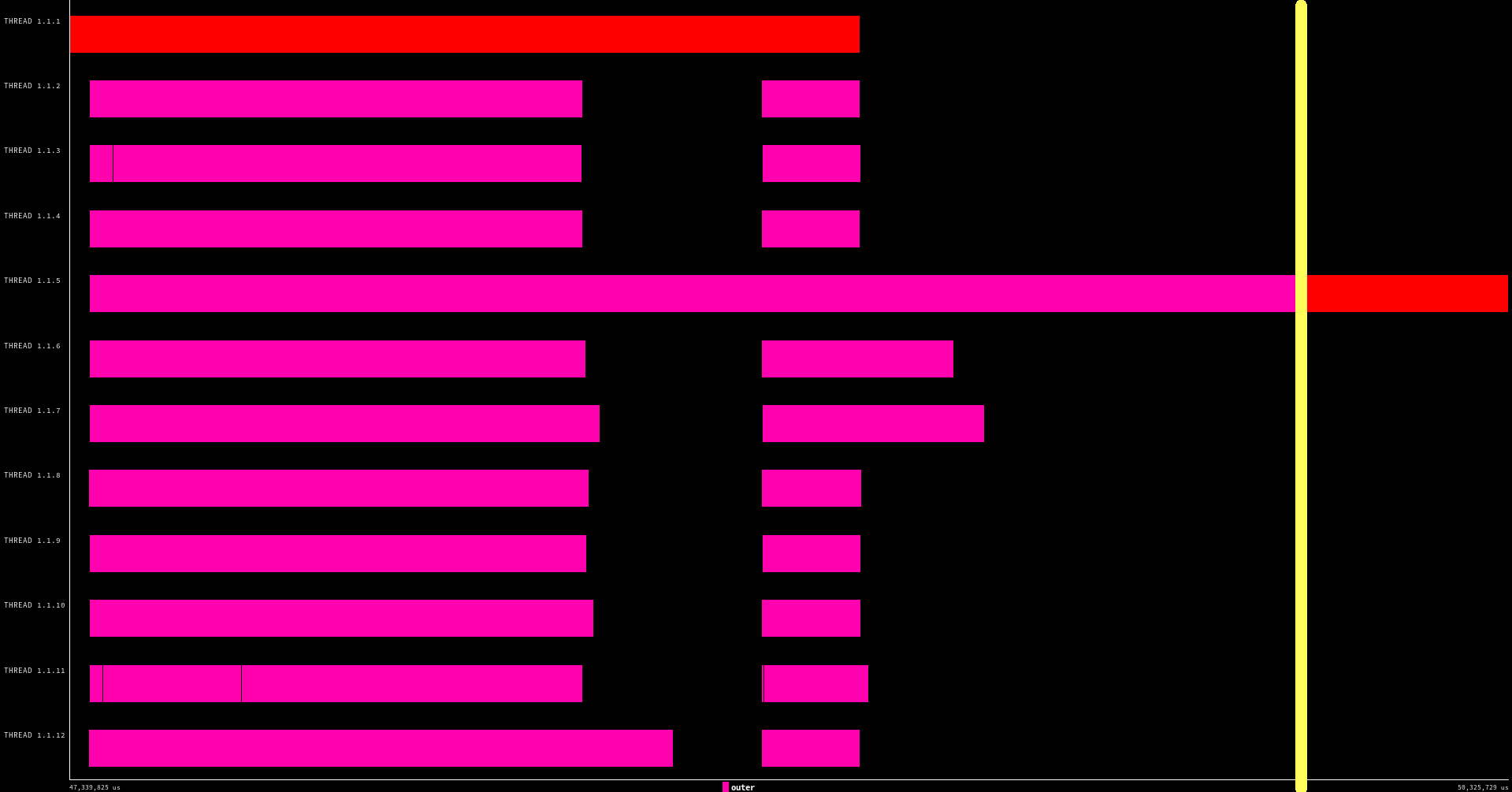}}
 \subfigure[With nested tasks (\NESTED)]{ \includegraphics[width=\linewidth,height=1.2cm]{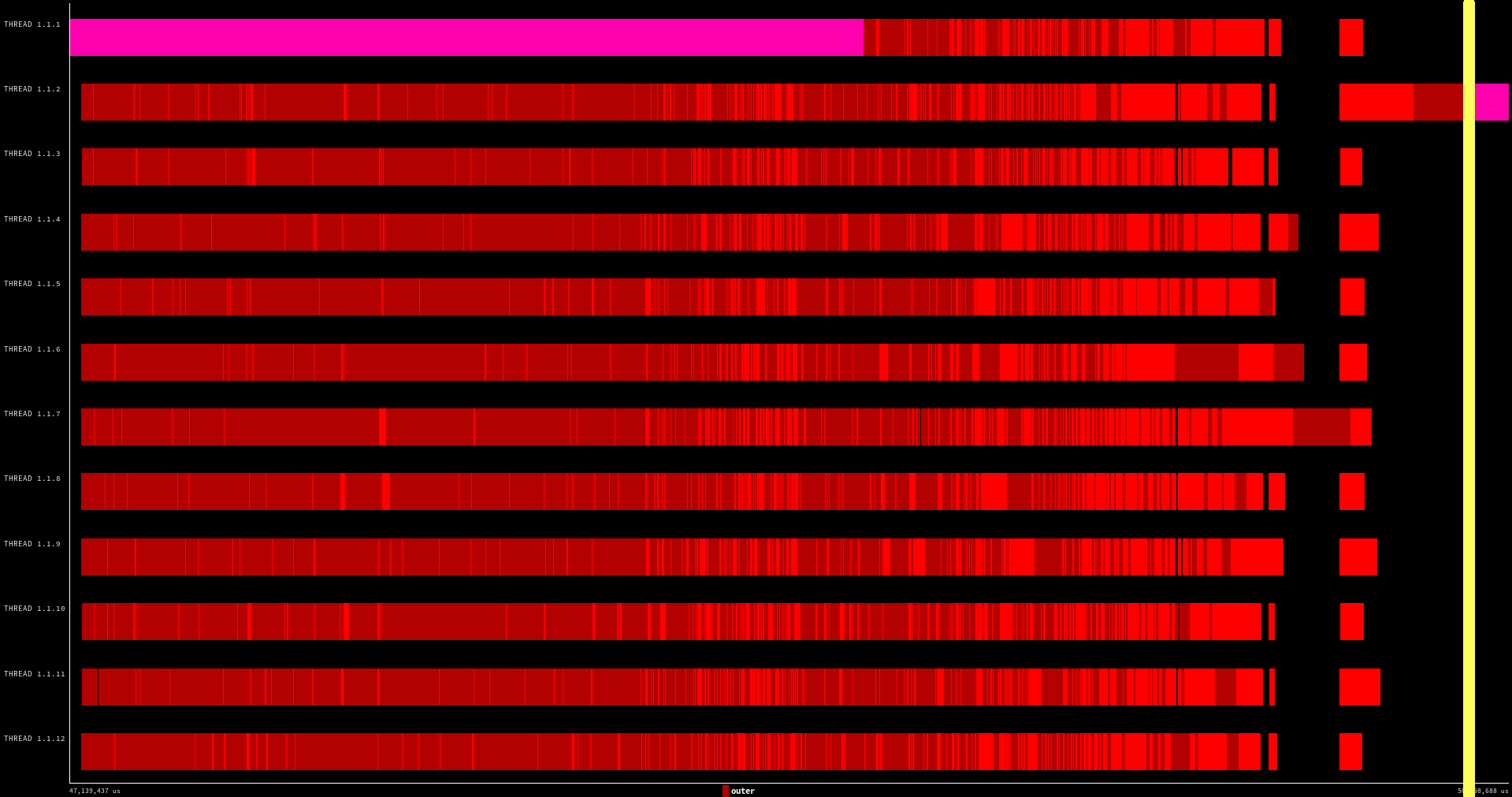}}
 \caption{Comparison of execution traces with or without nested tasks for \inline. The yellow line indicates the end of the parallel part.\label{fig.nesting_inline_1}}
\end{figure}

Figure~\ref{fig.nesting_bone010} shows the execution trace of Listing~\ref{algo.main} when \inner tasks are not instantiated (top, \NONESTED strategy), and when they are used (bottom, \NESTED strategy) for the input matrix
\bone. Our complete experimental setup is described in detail in Section~\ref{sec.expe.setup}.
The x-axis represents time while the y-axis represents different threads, 12 in this case. Each colored block corresponds to a task, 
and black blocks represent idle thread periods. The colors only indicate which part of the code is running. For \NONESTED, as there are only outer tasks, there is only one color (pink) for them. For \NESTED, the two shades of red indicate whether it
is an \outert (dark red) or \inner (light red) task.
When using nested tasks, one color is used for \outert tasks and another for \inner tasks.
As we can see, never using \inner tasks usually slows down the parallel run because it does not fully expose all sources of parallelism available. 
However, always using nesting can also slow down
the matrix factorization since task creation incurs overhead. 
Figure~\ref{fig.nesting_inline_1} shows such a case with the comparison of \NONESTED (top) and \NESTED (bottom) strategies for the input matrix \inline.
When too many \inner tasks are created, the main thread is fully dedicated to creating tasks as fast as possible and more synchronization is needed, and this becomes a
bottleneck for performance.
As an example with matrix \boneS, when we do not use nesting, we have 53,030 tasks and the ratio between the total time spent in task creation, scheduling and synchronization over the total time spent computing is around 11\%. Mainly one thread is responsible for managing the dependencies.
When using nesting, we reach 248,510 tasks and the same ratio goes up to 28\% and all the threads are involved into task management. The total computing time is also increased in that case.
Besides the task creation overhead problem, instantiating many tasks significantly increases the application memory footprint and sometimes overwhelms the system memory storage capacity, as Section~\ref{sec.expe} indicates.

For these reasons, we introduce \textit{selective nesting}. As demonstrated by the previous examples, there is a performance trade-off between exploiting the parallelism as much as possible
and the overhead of task creation. We need to determine what should be the best granularity for the tasks. Our idea is to select which \outert tasks will be split in several \inner tasks and which \outert tasks will not.
This means that, either all \inner tasks inside an \outert task will be used, or none of them.
To select which tasks we split, we introduce a threshold value, named $D$, on the number of updates received by each supernode from the previous ones. Each update corresponds
to one iteration of the inner loop, Line~\ref{line.inner_loop} in Listing~\ref{algo.main}: if the number of updates is higher than or equal to $D$ then the \outert task is split.
Otherwise, if there are not enough updates, all computation is kept embedded in the \outert task.
Thus, by varying $D$, we have a spectrum of heuristics to perform selective nesting with the two extreme cases being \NONESTED (when $D=\infty$) and \NESTED (when $D=1$).

To implement selective nesting, we use the \texttt{final} clause of OpenMP~\cite{openmp} in the definition of the \outert task:
if the clause evaluates to True, then the task cannot have nested tasks and all task directives inside of that task are ignored.
If the clause evaluates to False, then all task directives encountered inside the task are executed as usual. 

\subsection{The \OPT algorithm}
\label{sec.nesting.opt}

\begin{figure*}
\begin{center}
 \subfigure[\sdkq]{\includegraphics[width=0.32\linewidth]{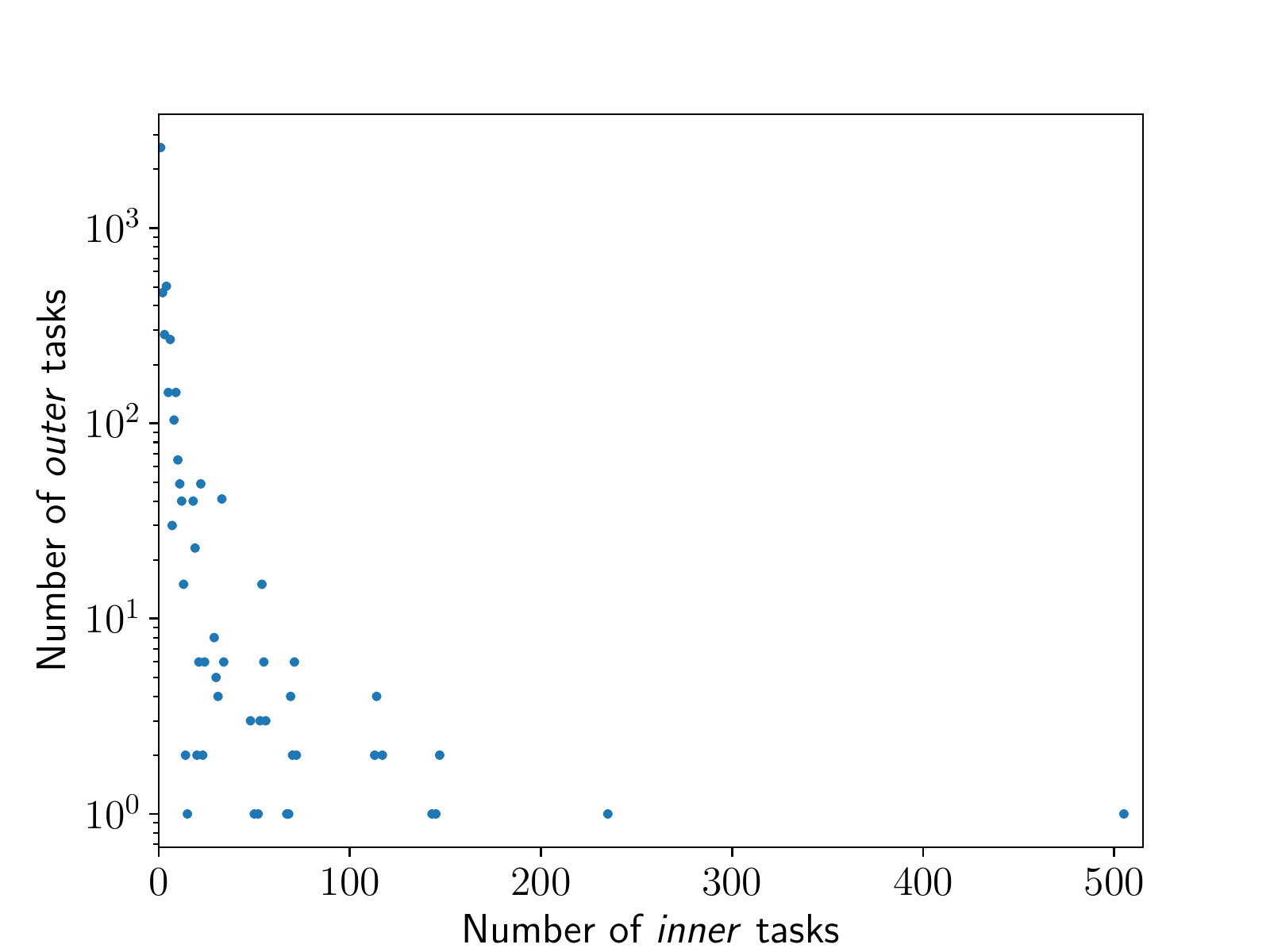}}
 \subfigure[\boneS]{\includegraphics[width=0.32\linewidth]{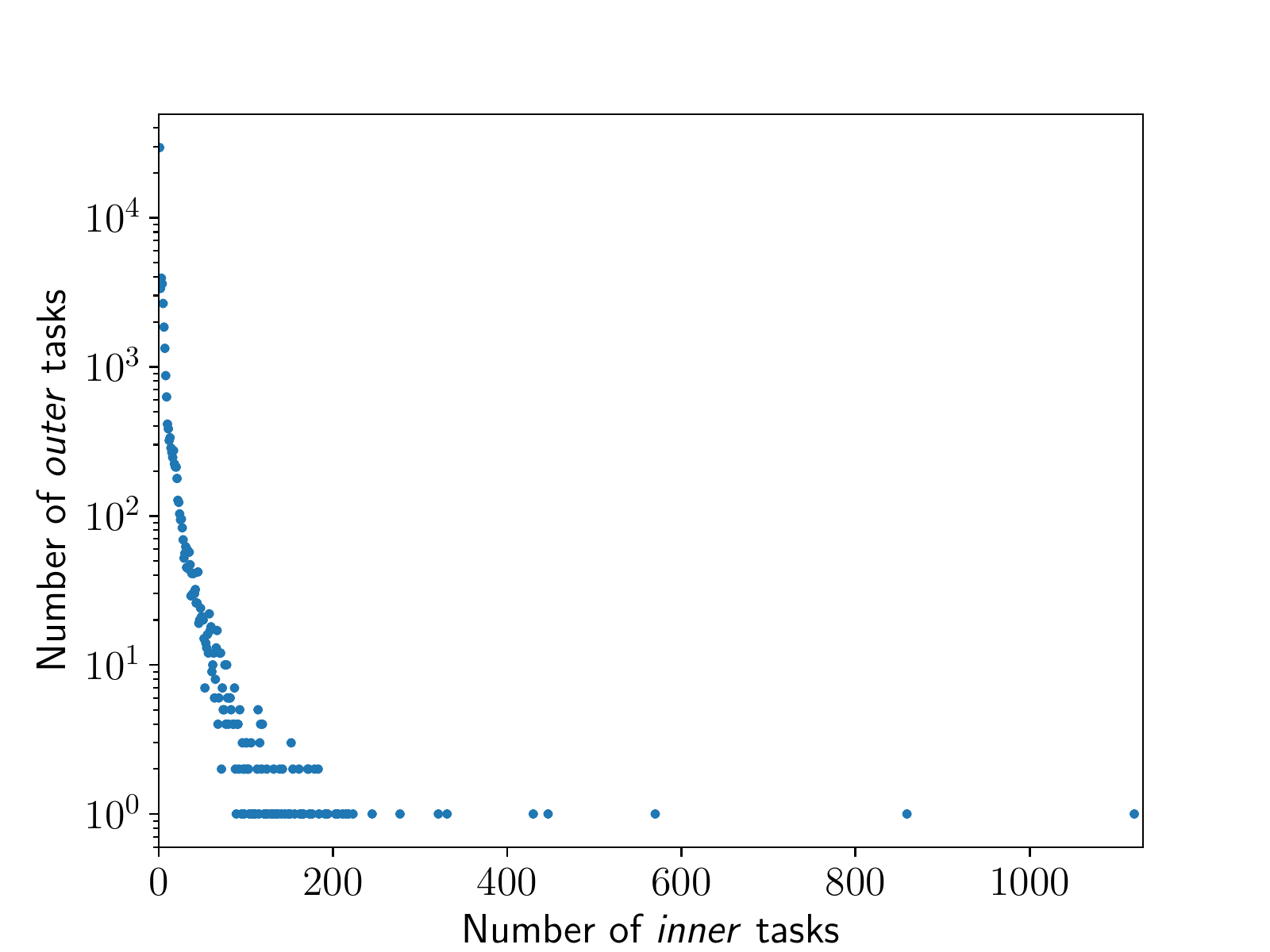}}
 \subfigure[\Gcircuit]{\includegraphics[width=0.32\linewidth]{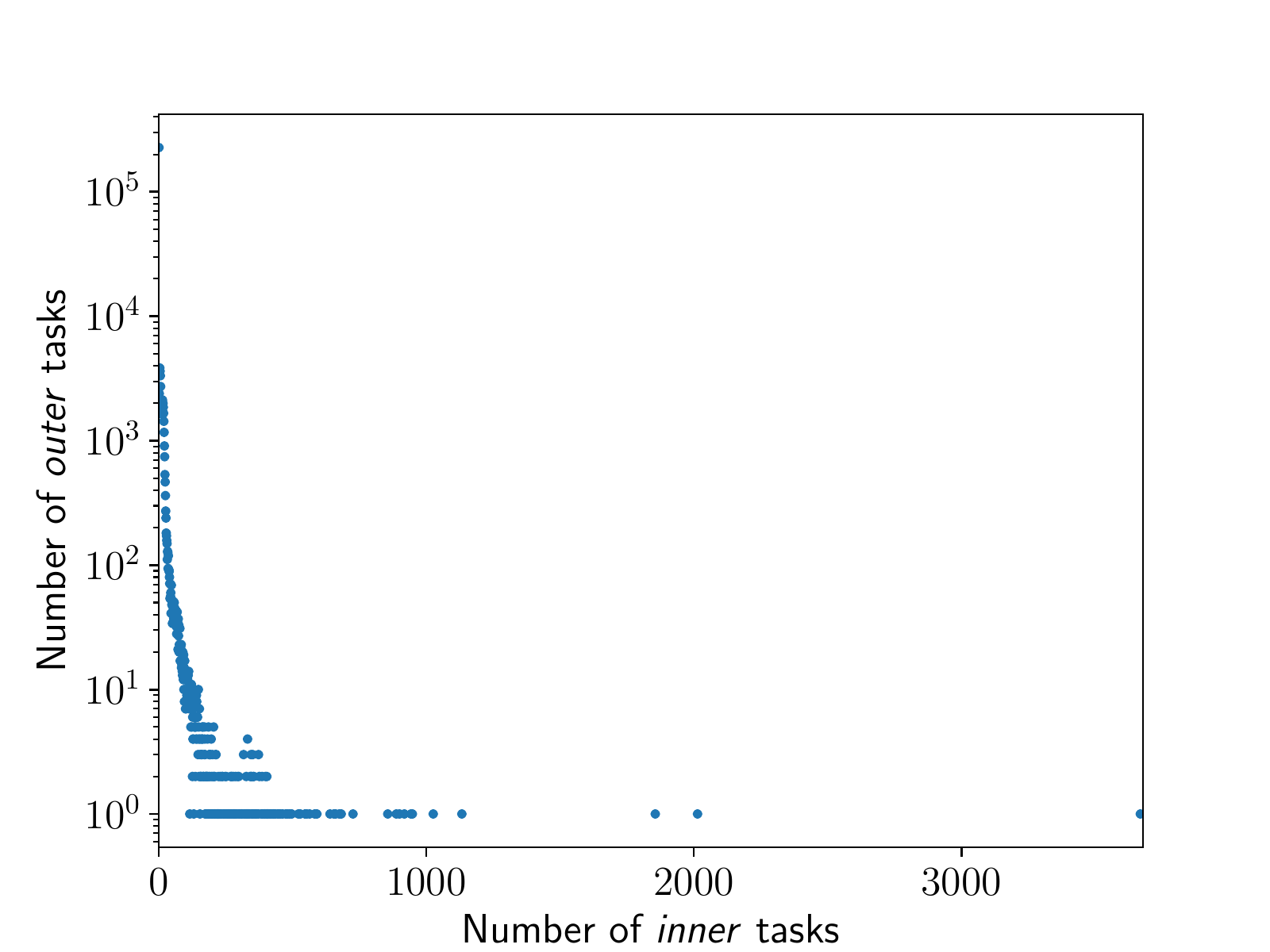}}
 \caption{Histogram of the number of \inner task per \outert task. Each point of x-axis value $i$ represents the number of \outert tasks that have $i$ \inner tasks.\label{fig.distrib}}
\end{center}
\end{figure*}

This section proposes the \OPT algorithm, a heuristic to drive selective nesting.
\OPT determines the best value of $D$ depending on the input matrix, instead of using a pre-determined value.
Indeed, setting a threshold of $D=100$ could be an efficient heuristic for some matrices, although it may happen that other $D$ values work better for some other sparse matrices.
To explain this variety of behaviors, Figure~\ref{fig.distrib} shows the distribution of \outert tasks in terms of the number of \inner tasks they instantiate
for three different matrices: \sdkq, \boneS and \Gcircuit. Each point of coordinates $(x,y)$ represents the number of \outert tasks ($y$) that have $x$ updates to perform, i.e. result in
the creation of $x$ \inner tasks if nesting is used.
Section~\ref{sec.expe.setup} describes our experimental setup.

The three distributions have the same profile: there is a huge number of \outert tasks that have only a few \inner tasks, and there are only a few \outert tasks with numerous \inner tasks.
The same shape of distribution can also be found for all the other sparse matrices we consider.
However, there is one impact aspect that changes across different matrices: the scale. 
For \sdkq the maximum number of \inner tasks is 505, while it is 1120 for \boneS, and 3669 for \Gcircuit. 
This means that setting a threshold of $D=300$ for \Gcircuit will create a much bigger proportion
of possible \inner tasks than a threshold of 300 for \sdkq. 
It appears natural to adapt the $D$ threshold to each input matrix instead of using a predefined one.

\begin{figure*}
\begin{center}
 \subfigure[\sdkq]{\includegraphics[width=0.32\linewidth]{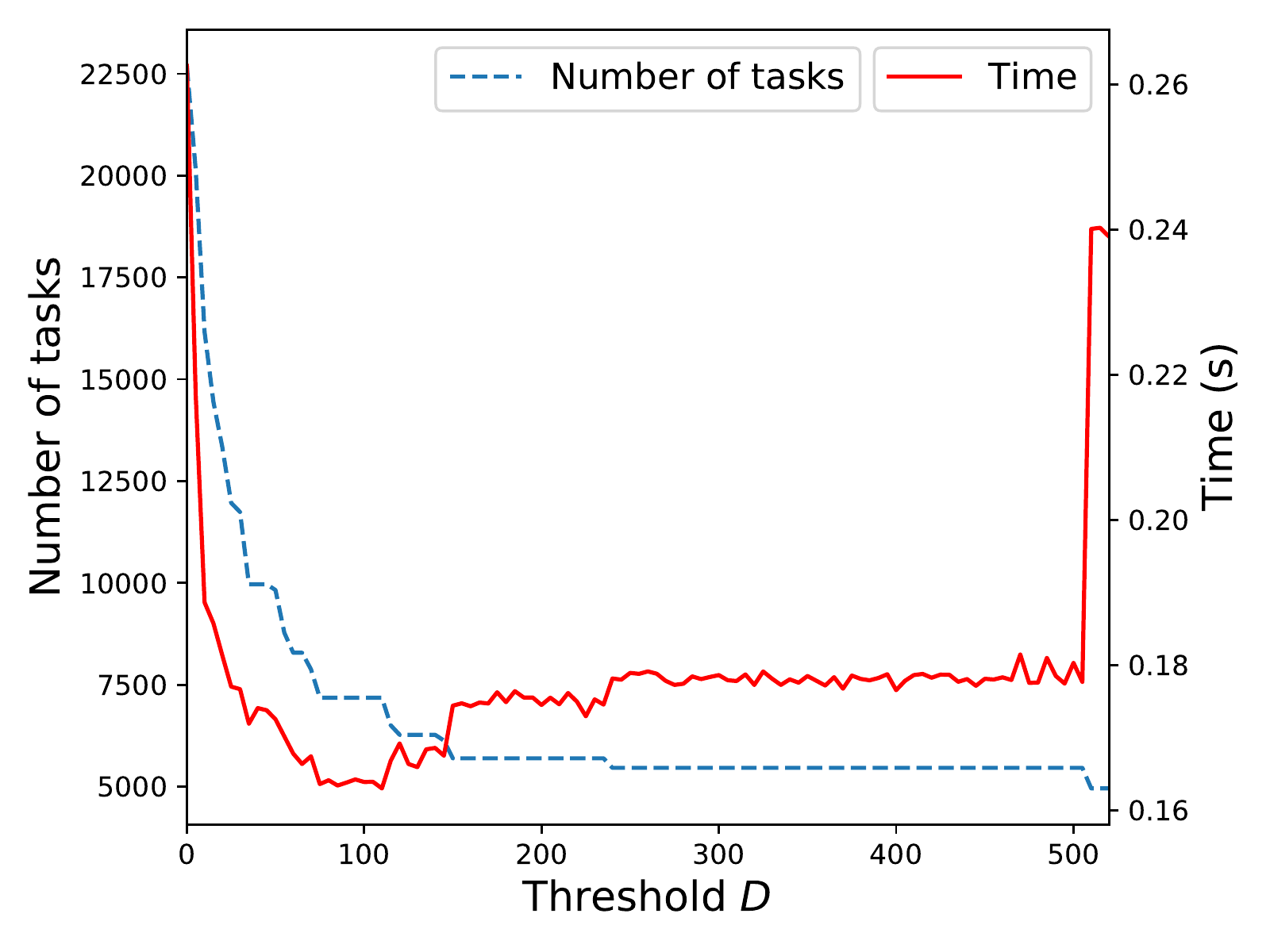}}
 \subfigure[\boneS]{\includegraphics[width=0.32\linewidth]{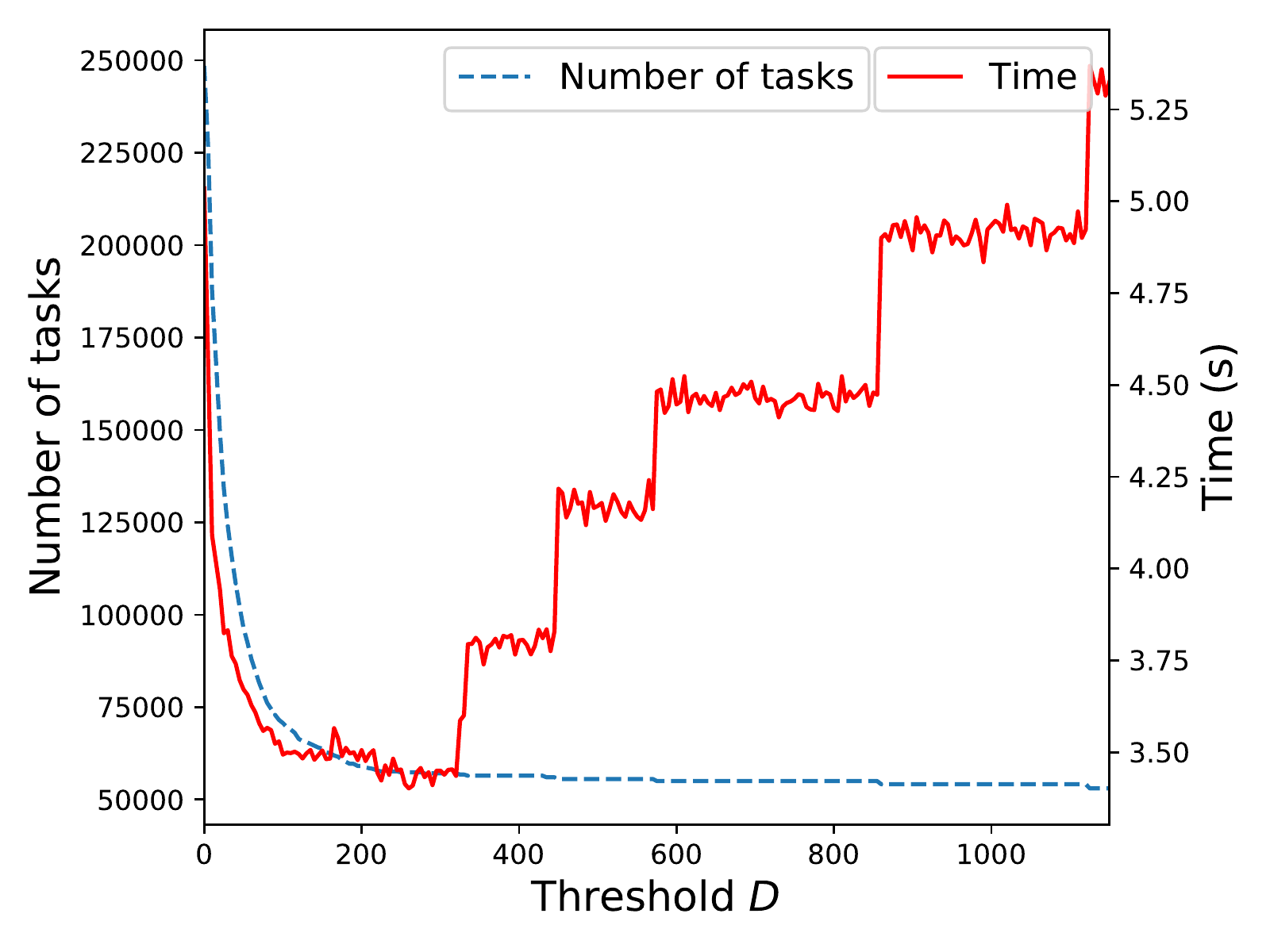}}
 \subfigure[\Gcircuit]{\includegraphics[width=0.32\linewidth]{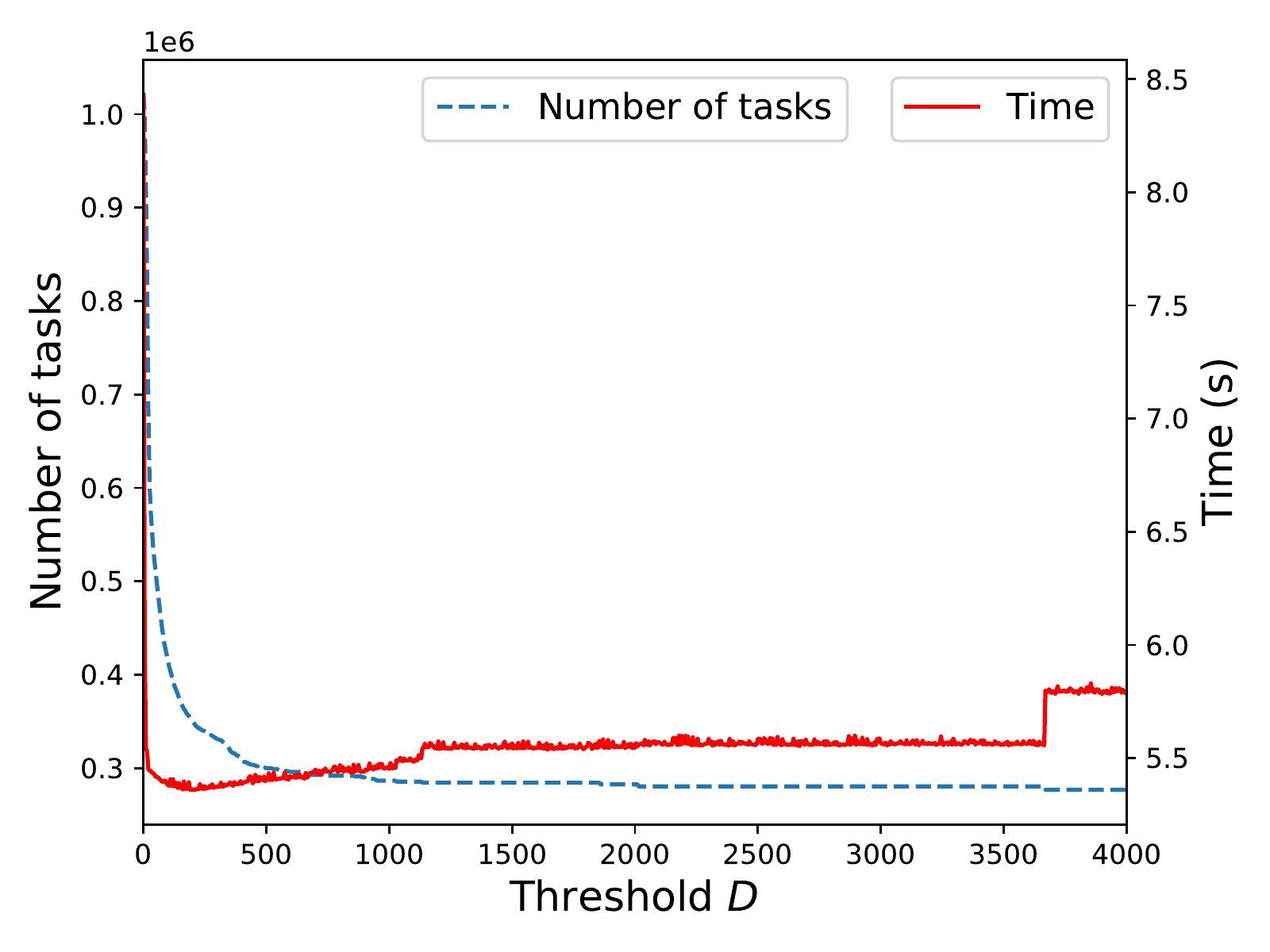}}
 \caption{Impact of $D$ on the execution time and the number of tasks for three input matrices.\label{fig.desc}}
\end{center}
\end{figure*}

We show the impact of $D$ on the Cholesky execution time for the three same matrices,
\sdkq, \boneS, and \Gcircuit, in Figure~\ref{fig.desc}.
We also report the corresponding number of tasks (\inner and \outert) created. 
The experiment considers 12 threads on an A64FX processor. Our experimental setup is described in detail in Section~\ref{sec.expe.setup}.
We can distinguish two different phases: First, when $D$ is small the execution time is large and decreases steadily as $D$ increases. 
Then,
when $D$ becomes larger, the execution time starts increasing again by jumps.
The first phase behavior, characterized by small values of $D$ and progressively better performance as $D$ value increases, can be explained by the
overhead of task creation and destruction, which is not negligible when a large amount of tasks is created. 
This
is the case when $D$ is small because most \outert tasks instantiate many \inner tasks. 
For the case of \boneS for example, the number of tasks is close to 250,000 when $D$ is 1, whereas
it is only around 55,000 when $D=400$. 
The shape of the curve for execution times actually follows that of the number of tasks in that first phase, since task instantiation is the bottleneck.
Therefore, the parallel execution is slowed down by the task creation overhead.

The second phase comes from the opposite behavior: task instantiation is not anymore the bottleneck and the jumps we can see correspond to the time taken by a long \outert task that does not instantiate 
\inner tasks as $D$ increases. 
The more we increase $D$, the longer the execution time becomes since there are more and more long tasks executing on one thread while the other threads are idle. 
The limit case corresponds the \NONESTED strategy.
For \Gcircuit, the shape is a bit different but we can still distinguish the two phases. For this matrix, the number of supernodes (hence the number of \outert tasks) is huge compared to other matrices. 
As a comparison, even without nesting, the \Gcircuit matrix creates more than 276,000 tasks. 
When too much nesting is introduced, i. e. $D$ is close to 1, the number of tasks
skyrockets and so does the execution time, which indicates that task creation is a serious performance bottleneck.

We design the \OPT algorithm to automatically find a good $D$ value for each input matrix.
Algorithm~\ref{algo.opt} shows a high-level description of \OPT.
As input parameters it uses the size of the matrix $n$, the number of supernodes $n_{super}$ and
the number of potential \inner tasks of an \outert task (array $C$), which CHOLMOD computes during the analysis phase.
\begin{algorithm}
\SetKwData{MC}{maxChildren}
\SetKwData{NT}{numTasks}
\SetKwData{NO}{numOuterTasks}
\SetKwData{GT}{goalTasks}
\SetKwInOut{Input}{input} \SetKwInOut{Output}{output}

\Input{$n$: size of the matrix\\$n_{super}$: number of supernodes\\$C$: array of size $n_{super}$ with the number of \inner tasks of each \outert task}
\Output{$D$}
 \GT $\leftarrow$ max($1.1\cdot n_{super}$,$\frac{n}{14}$)\;\label{line.step1}
 \MC $\leftarrow$ 0\;\label{line.step2.begin}
 \For{$i \leftarrow 1$ \KwTo $n_{super}$}{
   \MC $\leftarrow$  max(\MC,$C[i]$)\;
 }
 Create array $T$ of size \MC initialized at 0\;
 \For{$i \leftarrow 1$ \KwTo $n_{super}$}{
   $T[C[i]] \leftarrow T[C[i]]+1$\;
 }\label{line.step2.end}
 $D \leftarrow $ \MC$+~1$\;\label{line.step3.begin}
 \NO $\leftarrow$ 0\;
 \NT $\leftarrow n_{super}$\;
 \While{ (\NT $<$ \GT \textbf{or} $D > 0.3\cdot$\MC \textbf{or} \NO $< \frac{n_{super}}{1000}$) \textbf{and} $D>0$ }{
    $D \leftarrow D-1$\;
    \NO $\leftarrow$ \NO$+~T[D]$\;
    \NT $\leftarrow$ \NT$+~D\cdot T[D]$\;
 }
 \Return{$D$}\;\label{line.step3.end}
 \caption{\OPT.\label{algo.opt}}
\end{algorithm}

The first step of \OPT is to determine a reasonable number of tasks (variable \texttt{goalTasks}), which is done at Line~\ref{line.step1}.
We experimentally determine the values of $D$ that provide the lowest execution time for several matrices (including the ones presented in Figure~\ref{fig.desc}) in order to find
correlations between the different matrices.
The behavior of these relevant sparse matrices indicates that allowing the ratio between the size of the matrix and the number of created tasks to be above 14 implies a too large $D$ value. 
Just slightly overestimating $D$ can lead
to a significant increase in the execution time, as execution time sudden jumps in Figure~\ref{fig.desc} indicate. 
Thus, we target
the smallest number of tasks which reduces this ratio just below 14.
However, if the average size, in terms of number of columns, of each supernode was already below 14, the minimum number of tasks would be targeted, i.e. no nesting, according to this criteria.
However, our data show that, even in this case, allowing some nesting reduces the execution time, so we require that at least 10\% more tasks are created.
We thus take the maximum between $\frac{n}{14}$ and $1.1\cdot n_{super}$ as the minimum number of tasks to reach, with $n$ the size of the matrix, and $n_{super}$ the number of supernodes.

The second step is to sort the \outert tasks by decreasing number of \inner tasks. 
This step is done through the application of bucket sort to the \texttt{C} array in the \texttt{T} array, through Lines~\ref{line.step2.begin} to~\ref{line.step2.end}.
Once the bucket sort is done, \texttt{T[i]} contains the number of \outert tasks that have $i$ updates, i.e. \texttt{T} contains the information presented
in Figure~\ref{fig.distrib}.

The third step of the algorithm is to determine $D$ by going through the \texttt{T} array backwards, which corresponds to Lines~\ref{line.step3.begin} to~\ref{line.step3.end}. 
Additionally, we enforce $D$ to be smaller than $30\%$ of the maximum number of \inner tasks in an \outert task, as our preliminary analysis on the optimal $D$ value for a set of relevant sparse matrices indicates.
We also ensure that at least the largest 0.1\% \outert tasks always instantiate other \inner tasks. 
This takes into account the case when the largest \outert tasks (in terms of number of \inner tasks)
account for more than 10\% of new created tasks. 
If this occurs, only the largest tasks are split, limiting the parallelism for the supernodes that are factorized first, as big \outert tasks tend to be executed at the end
due to the ordering in the \textit{analyze} phase.

As \texttt{maxChildren} is in practice much smaller than the number of supernodes, the algorithm runs in $O(n_{super})$ where $n_{super}$ is the number of supernodes.
Its cost is thus negligible in front of the factorization time.

\subsection{The \OPTIF algorithm}
\label{sec.nesting.optif}

This Section presents the \OPTIF algorithm, which is an extension of \OPT. 
\OPT determines a threshold $D$ based on the estimated size of each \outert task. 
If an \outert task has more than $D$ (as computed by \OPT) \inner tasks, then all the \inner tasks are created.
However, if an \inner task embeds a tiny amount of computation, it might be worth keeping it in the \outert task to avoid paying
the cost of task creation, which would be of the same order of magnitude.

To deal with this problem we introduce an additional threshold to drive \inner task creation. 
An \inner task is composed of one SYRK kernel and one GEMM kernel, so we estimate its cost in terms of floating-point operations using previously proposed methods based on computational complexities~\cite{trefethen97}.
If the cost of the \inner task is smaller than the threshold, the task will not be created and computation will be embedded in the original \outert task. 
Otherwise, the task is created as usual.
We experimentally tuned the threshold to be 50,000 floating-point operations and we refer to this algorithm as \OPTIF. 
OpenMP does not provide any way of controlling the task creation of a single nested task based on a condition\footnote{The \textbf{if} clause only allows to defer the execution of a task~\cite{ompss,openmp}.}, so the implementation was done by
having the code of the \inner tasks preceded by an OpenMP \texttt{task} directive guarded by an if-then construct. 

\subsection{Additional Methods to Exploit Parallelism Within Supernodes}
\label{sec.nesting.blas}
While Sections~\ref{sec.nesting.opt} and~\ref{sec.nesting.optif} describe methods to adjust nested tasking within supernodes without exploiting any parallelism within BLAS calls,
this section describes an additional method to exploit parallelism within supernodes based on multi-threaded BLAS kernels.
%
%
This method sequentially factorizes supernodes via BLAS/LAPACK calls (POTRF, TRSM, SYRK and GEMM) that run on multiple threads.
These multi-threaded BLAS kernels are very efficient for dense matrices, which makes them more suitable for large and dense supernodes as they require larger updates and factorizations.
We refer to this method based on multi-threaded BLAS kernels as \MTBLAS.
%
To improve our algorithms, we define a hybrid approach that switches between \OPTIF (or \OPT) and \MTBLAS.
This hybrid algorithm can be summarized as follows:
\begin{compactitem}
\item if the average supernode size is greater than 50: use \MTBLAS;
\item if the average supernode size is greater than 20 and the density of the matrix, i.e. the number of non-zeros divided by the total number of cells, is below $10^{-4}$: use \MTBLAS;
\item otherwise, use \OPTIF (or \OPT) with sequential BLAS calls and OpenMP task directives.
\end{compactitem}

We will compare our hybrid approach to \MTBLAS in Section~\ref{sec.expe}. For simplicity, we will refer to
this hybrid approach as \OPTIF (or \OPT) through the end of the paper, even though depending on the input matrix,
the algorithm may rely on \MTBLAS.

%% file: expe_arxiv.tex
This section is dedicated to the performance evaluation of all the heuristics described in Section~\ref{sec.nesting}.
Section~\ref{sec.expe.setup} describes in detail our experimental setup, and Section~\ref{sec.expe.results1} compares
the performance of \OPT and \OPTIF with \MTBLAS on a wide range of matrices.
Section~\ref{sec.expe.results1} also considers two more approaches: \NONESTED, which never creates \inner tasks, and \NESTED which always does so.

\subsection{Experimental setup}
\label{sec.expe.setup}

This subsection describes the hardware platform and the software stack we use in our experimental campaign.
We consider a 4-socket A64FX~\cite{A64FX} machine, which implements the Armv8.2 architecture and supports SVE instructions:
%
%
\begin{itemize}
 \item 4 sockets of 12-core A64FX CPU chips, each @2.20GHz for a total of 48 cores per node
 \item 32 GB of main memory HBM2
 \item TofuD network
\end{itemize}

We use the Red Hat Enterprise Linux Server 8.1 operating system, Fujitsu optimized BLAS kernels, and the vendor-optimized fcc 4.2.0b compiler.
To support the execution of our OpenMP codes using task directives we use the OmpSs-1 19.06 as a runtime system~\cite{ompss}. 
This runtime system fully supports OpenMP and makes use of nanos++ and mercurium as runtime library and compiler.
Our results consider hybrid versions of the \OPT and \OPTIF algorithms that use either multi-threaded BLAS kernels or OpenMP tasks.
We report the performance that these hybrid versions achieve based on the condition that selects using either multi-threaded BLAS kernels or OpenMP tasks.

\begin{table}
 
\begin{minipage}{0.49\linewidth}
\caption{Matrices of Group 1.\label{table.group1}}
\resizebox{\linewidth}{!}{
\begin{tabular}{|cccc|}
\hline
Matrix Name & Number of rows & Non-zeros & Problem type \\
\hline
 \hline
\texttt{bcsstk34} & 588 & 21418 & Structural \\ 
 \hline
\texttt{msc01050} & 1050 & 26198 & Structural \\ 
 \hline
\texttt{bcsstk21} & 3600 & 26600 & Structural \\ 
 \hline
\texttt{plbuckle} & 1282 & 30644 & Structural \\ 
 \hline
\texttt{plat1919} & 1919 & 32399 & 2D/3D \\ 
 \hline
\texttt{bcsstk11} & 1473 & 23241 & Structural \\ 
 \hline
\texttt{msc00726} & 726 & 34518 & Structural \\ 
 \hline
\texttt{nasa1824} & 1824 & 39208 & Structural \\ 
 \hline
\texttt{Trefethen\_2000} & 2000 & 41906 & Combinatorial \\ 
 \hline
\texttt{msc01440} & 1440 & 44998 & Structural \\ 
 \hline
\texttt{bcsstk23} & 3134 & 45178 & Structural \\ 
 \hline
\end{tabular}
}
\end{minipage} 
\begin{minipage}{0.49\linewidth}
\caption{Matrices of Group 2.\label{table.group2}}
\resizebox{\linewidth}{!}{
\begin{tabular}{|cccc|}
\hline
Matrix Name & Number of rows & Non-zeros & Problem type \\
\hline
\hline
\texttt{  nasa4704} & 4704 & 104756 & Structural \\ 
 \hline
\texttt{  crystm01} & 4875 & 105339 & Materials \\ 
 \hline
\texttt{  bcsstk15} & 3948 & 117816 & Structural \\ 
 \hline
\texttt{  bodyy4} & 17546 & 121550 & Structural \\ 
 \hline
\texttt{  aft01} & 8205 & 125567 & Acoustics \\ 
 \hline
\texttt{  bodyy5} & 18589 & 128853 & Structural \\ 
 \hline
\texttt{  bodyy6} & 19366 & 134208 & Structural \\ 
 \hline
\texttt{  bcsstk18} & 11948 & 149090 & Structural \\ 
 \hline
\texttt{  bcsstk24} & 3562 & 159910 & Structural \\ 
 \hline
\texttt{  Muu} & 7102 & 170134 & Structural \\ 
 \hline
\texttt{  nasa2910} & 2910 & 174296 & Structural \\ 
 \hline
\texttt{  t2dah\_e} & 11445 & 176117 & Duplicate Model Reduction \\ 
 \hline
\texttt{  obstclae} & 40000 & 197608 & Optimization \\ 
 \hline
\texttt{  jnlbrng1} & 40000 & 199200 & Optimization \\ 
 \hline
\end{tabular}
}
\end{minipage}
\end{table}

\begin{table}
\begin{minipage}{0.49\linewidth}
\caption{Matrices of Group 3.\label{table.group3}}
\resizebox{\linewidth}{!}{
\begin{tabular}{|cccc|}
\hline
Matrix Name & Number of rows & Non-zeros & Problem type \\
\hline
\hline
\texttt{cfd2} & 123440 & 3085406 & Computational Fluid Dynamics \\ 
 \hline
\texttt{nd3k} & 9000 & 3279690 & 2D/3D \\ 
 \hline
\texttt{shipsec8} & 114919 & 3303553 & Structural \\ 
 \hline
\texttt{shipsec1} & 140874 & 3568176 & Structural \\ 
 \hline
\texttt{Dubcova3} & 146689 & 3636643 & 2D/3D \\ 
 \hline
\texttt{parabolic\_fem} & 525825 & 3674625 & Computational Fluid Dynamics \\ 
 \hline
\texttt{s3dkt3m2} & 90449 & 3686223 & Structural \\ 
 \hline
\texttt{smt} & 25710 & 3749582 & Structural \\ 
 \hline
\texttt{ship\_003} & 121728 & 3777036 & Structural \\ 
 \hline
\texttt{ship\_001} & 34920 & 3896496 & Structural \\ 
 \hline
\texttt{cant} & 62451 & 4007383 & 2D/3D \\ 
 \hline
\texttt{offshore} & 259789 & 4242673 & Electromagnetics \\ 
 \hline
\texttt{pdb1HYS} & 36417 & 4344765 & Weighted Unidrected Graph \\ 
 \hline
\texttt{s3dkq4m2} & 90449 & 4427725 & Structural \\ 
 \hline
\texttt{thread} & 29736 & 4444880 & Structural \\ 
 \hline
\texttt{shipsec5} & 179860 & 4598604 & Structural \\ 
 \hline
\texttt{consph} & 83334 & 6010480 & 2D/3D \\ 
 \hline
\end{tabular}
}
\end{minipage}
\begin{minipage}{0.49\linewidth}
\caption{Matrices of Group 4.\label{table.group4}}
\resizebox{\linewidth}{!}{
\begin{tabular}{|cccc|}
\hline
Matrix Name & Number of rows & Non-zeros & Problem type \\
\hline
\hline
\texttt{apache2} & 715176 & 4817870 & Structural \\
\hline
\texttt{ecology2} & 999999 & 4995991 & 2D/3D \\ 
 \hline
\texttt{tmt\_sym} & 726713 & 5080961 & Electromagnetics \\ 
 \hline
\texttt{boneS01} & 127224 & 5516602 & Model Reduction \\
\hline
\texttt{G3\_circuit} & 1585478 & 7660826 & Circuit Simulation \\
\hline
\texttt{thermal2} & 1228045 & 8580313 & Thermal \\
\hline
\texttt{af\_shell3} & 504855 & 17562051 & Subsequent Structural \\
\hline
\texttt{StocF-1465} & 1465137 & 21005389 & Computational Fluid Dynamics \\
\hline
\texttt{Fault\_639} & 638802 & 27245944 & Structural \\
\hline
\texttt{nd24k} & 72000 & 28715634 & 2D/3D \\
\hline
\texttt{inline\_1} & 503712 & 36816170 & Structural \\
\hline
\texttt{Emilia\_923} & 923136 & 40373538 & Structural \\
\hline
\texttt{boneS10} & 914898 & 40878708 & Model Reduction \\
\hline
\texttt{ldoor} & 952203 & 42493817 & Structural \\
\hline
\texttt{bone010} & 986703 & 47851783 & Model Reduction \\
\hline
\texttt{Hook\_1498} & 1498023 & 59374451 & Structural \\
\hline
\texttt{audikw\_1} & 943695 & 77651847 & Structural \\
\hline
\texttt{Flan\_1565} & 1564794 & 114165372 & Structural \\
 \hline
\end{tabular}
}
\end{minipage}
\end{table}

We evaluate the different algorithms considering 60 input matrices from the SuiteSparse Matrix Collection~\cite{suitesparseMC}. 
We define 4 different groups of symmetric definite positive matrices based on their number of non-zero elements:
 \begin{itemize}
  \item Group 1 is described in Table~\ref{table.group1}. It contains matrices from 10,000 to 50,000 non-zero elements. 
  \item Group 2 is described in Table~\ref{table.group2}. It contains matrices from 100,000 to 200,000 non-zero elements. 
  \item Group 3 is described in Table~\ref{table.group3}. It contains matrices from 3,000,000 to 6,000,000 non-zero elements.
  \item Group 4 is described in Table~\ref{table.group4}. It contains matrices with at least 4,800,000 non-zero elements, and they are among the largest matrices in terms of number of rows in the SuiteSparse collection.
 \end{itemize}

The experiments report the average over 100 runs for matrices of Groups 1 and 2, and over 10 runs for matrices of Groups 3 and 4.

\subsection{Performance evaluation of \OPT and \OPTIF}
\label{sec.expe.results1}


\begin{figure*}
 \includegraphics[width=\linewidth]{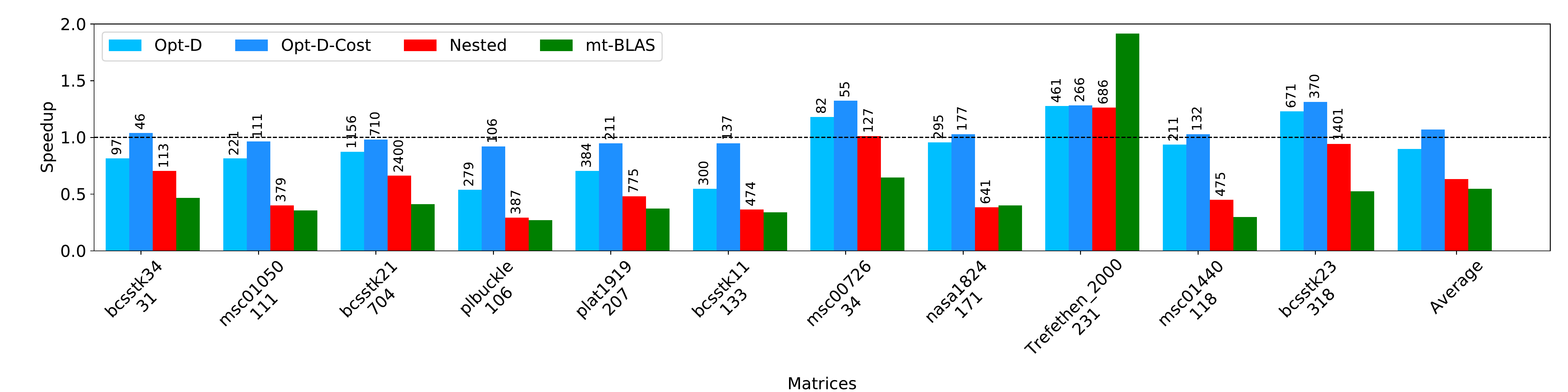}
 \caption{Speed-up relative to \NONESTED of 4 algorithms on \ARM using 12 threads for matrices of Group 1.\label{fig.eval.g1.S}}
\end{figure*}

\begin{figure*}
 \includegraphics[width=\linewidth]{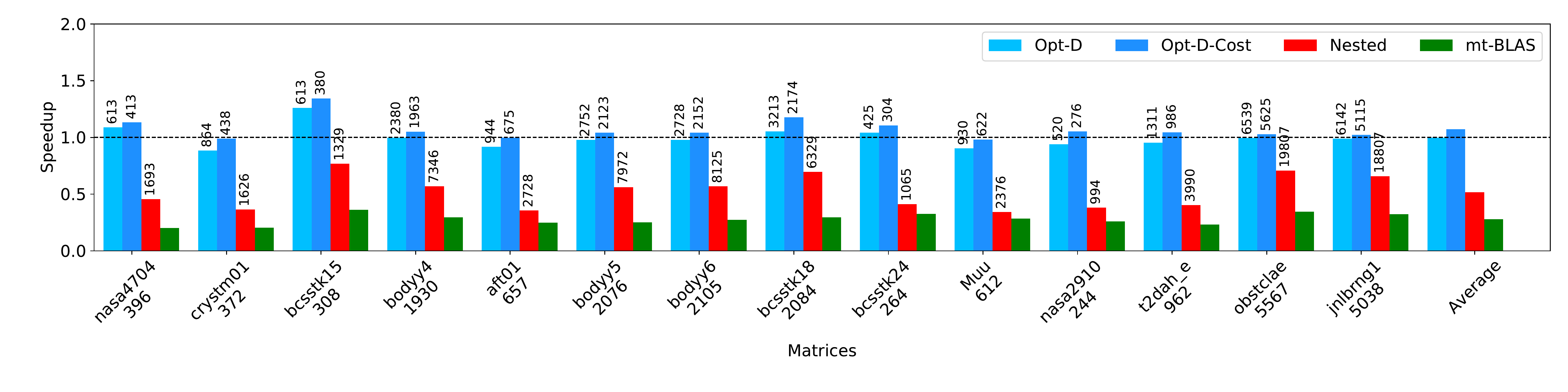}
 \caption{Speed-up relative to \NONESTED of 4 algorithms on \ARM using 12 threads for matrices of Group 2.\label{fig.eval.g2.S}}
\end{figure*}

\begin{figure*}
 \includegraphics[width=\linewidth]{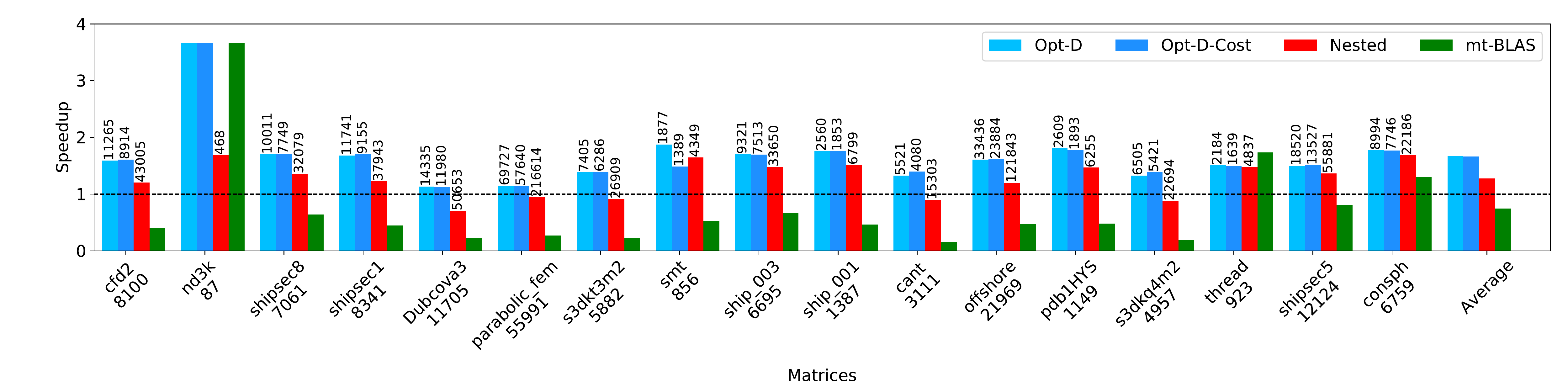}
 \caption{Speed-up relative to \NONESTED of 4 algorithms on \ARM using 12 threads for matrices of Group 3.\label{fig.eval.g3.S}}
\end{figure*}

\begin{figure*}
 \includegraphics[width=\linewidth]{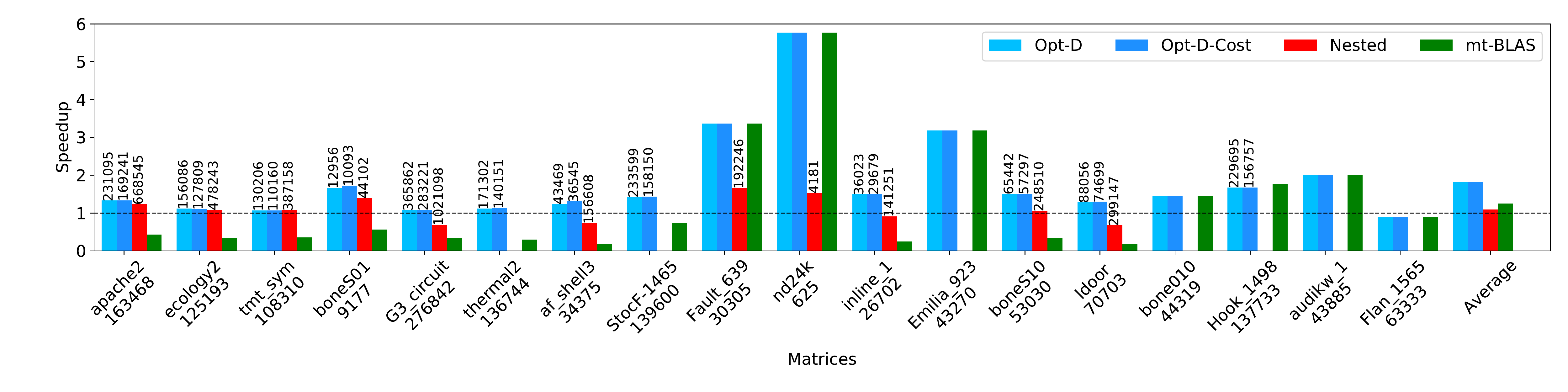}
 \caption{Speed-up relative to \NONESTED of 4 algorithms on \ARM using 12 threads for matrices of Group 4.\label{fig.eval.g4.S}}
\end{figure*}

This section compares the performance of \OPT, \OPTIF, \NESTED, \NONESTED and \MTBLAS.
Figures~\ref{fig.eval.g1.S},~\ref{fig.eval.g2.S},~\ref{fig.eval.g3.S}, and~\ref{fig.eval.g4.S} show the speed-up of all algorithms with respect to \NONESTED for group matrices 1, 2, 3, and 4, respectively. 
We report the average
speed-ups for each group on the right-hand side of each plot.
On top of each bar, we also indicate the total number of tasks executed for each algorithm. The number of tasks executed using \NONESTED, i.e. the number of supernodes for the matrix, is reported below the matrix name.
Note that for \MTBLAS implementation, even when used by \OPT and \OPTIF heuristics, we do not have any number of tasks as the parallelism comes from classical multi-threading.

Figure~\ref{fig.eval.g1.S} reports performance results when considering matrices belonging to Group~1.
\OPTIF is the only approach that provides better average performance than \NONESTED since it achieves a 1.07$\times$ average speedup.
Results of Figure~\ref{fig.eval.g1.S} indicate that \OPTIF provides very significant performance improvements with respect to \NONESTED for some matrices (e.g. 1.32$\times$ for the case of \texttt{msc00726}).
The average speed-ups of \OPT is 0.90$\times$ for Group~1 with respect to \NONESTED, which clearly indicates the benefits of adding the additional condition to drive nested parallelism that Section~\ref{sec.nesting.optif} introduces. 
We can also notice that in most cases \MTBLAS is much slower than \OPTIF, averaging
a speed-up of only 0.55$\times$. 
This is due to the size and sparsity of supernodes, which make BLAS kernels and \inner tasks very short. 
The parallelization
of the BLAS kernel then leads to poor performance, and the overhead due to creation of nested tasks is not negligible for the case of \OPT.
This can be clearly seen for matrix \texttt{bcsstk11}, where \OPTIF has a speed-up of 0.95$\times$, compared to 0.55$\times$ for \OPT, by removing the creation of small \inner tasks. \OPTIF was designed to counter this problem
and thus gives a good performance overall.

Figure~\ref{fig.eval.g2.S} reports the performance of all considered approaches on matrices belonging to Group~2.
\NESTED obtains worse performance than \NONESTED, a 0.52$\times$ average speed-up, while \MTBLAS only achieves a speed-up of 0.28$\times$ compared to \NONESTED.
The poor performance of these two techniques, \NESTED and \MTBLAS, is due to the sparsity and the small size of matrices supernodes.
In contrast, \OPTIF achieves an average speed-up of 1.07$\times$ for Group~2 with respect to \NONESTED.
Precisely, \OPTIF
is 3.84$\times$ faster than \MTBLAS. and 2.07$\times$ faster than \NESTED for Group~2.

For Group~3, on Figure~\ref{fig.eval.g3.S}, 
\MTBLAS has an average speed-up of 0.75$\times$, while \NESTED grants a 1.28$\times$ average speed-up with respect to \NONESTED. 
\OPT and \OPTIF overcome the other
approaches by reaching an average speed-up of 1.68$\times$ and 1.67$\times$ respectively. For Group~3, \OPTIF is thus 1.30$\times$ faster than \NESTED and 2.23$\times$ faster than \MTBLAS.
The performance of \OPT and \OPTIF is very similar for the 18 matrices while \MTBLAS
is clearly the worst heuristic, reaching a speed-up of only 0.15$\times$ for \texttt{cant}. 
Contrarily to Group~2, 
\NESTED
is a more efficient heuristic than \NONESTED for Group~3, which can be explained by two facts: First, \inner tasks embed a significant amount of computation. 
Second, when reaching the root of the elimination tree, tree parallelism experiences a significant reduction and \NONESTED induces long sequential tasks at the end of the execution since supernodes are large for Group~3 matrices.

There is one special matrix in Group~3: \texttt{nd3k}. This matrix is much denser than the others, has less rows and columns than the others and has
especially huge supernodes. While most matrices so far have an average supernode size between 5 and 25, that of \texttt{nd3k} is 103.45.
This matrix has a few large and dense supernodes. 
Large and dense supernodes mean
that multi-threaded BLAS kernels are efficient,which is why \MTBLAS gets a speed-up of 3.66$\times$ for \texttt{nd3k}, but so does our heuristics \OPT and \OPTIF
as the average supernode size is well above 50. 

Finally, we can derive the same conclusions for Group~4 as for Group~3. \OPT and \OPTIF are the best heuristics with average speed-ups of 1.82$\times$ and 1.83$\times$, respectively.
There are several matrices where \OPT and \OPTIF use \MTBLAS: \texttt{Fault\_639}, \texttt{Emilia\_923}, \texttt{bone010}, \texttt{audikw\_1}, \texttt{Flan\_1565} and \texttt{nd24k}.
The last one is a matrix with the same properties as \texttt{nd3k} of Group~3, while the five others have densities smaller than $10^{-3}$ and an average supernode size between 20 and 25.
Except for \texttt{Flan\_1565}, the speed-up reached by the use of multi-threaded BLAS kernels is greater than 1$\times$ for all this subset of matrices. 
We can clearly see the benefits or our hybrid approach for matrices such as \texttt{af\_shell3}, where the speed-up of \MTBLAS drops to only 0.19$\times$, while \OPTIF is at 1.31$\times$.
The average speed-up of \MTBLAS is 1.25$\times$, meaning that \OPTIF has a 1.46$\times$ speed-up with respect to \MTBLAS.
Generally, the speed-up of \OPT and \OPTIF is better than \NESTED, with
the advantage of consuming less memory. Indeed, the speed-up for \NESTED is not reported for all inputs because of the memory consumption of all instantiated tasks is above the machine capacity. 
The total memory capacity of the A64FX machine is not enough to handle \NESTED, at least for one of the ten runs, for the seven following matrices:
\thermal, \texttt{StocF-1465}, \texttt{Emilia\_923}, \bone, \audi, \texttt{Hook\_1498}, \texttt{Flan\_1565}.

To summarize, the improvement made with \OPTIF over \OPT is more noticeable for small matrices. 
It thus makes \OPTIF an efficient algorithm over a large number of different input matrices (a total of 60 in this
evaluation). 
Its performance is equal or better than \NONESTED, always better than \NESTED and is rarely beaten by \MTBLAS as we are able to detect
the matrices for which it is better not to rely on a task-based decomposition. Importantly, in each group, we can see that \MTBLAS can really slow down
the factorization phase for some matrices.
On average over the 60 matrices, \MTBLAS, \OPT and \OPTIF achieve a speed-up of 0.75$\times$, 1.42$\times$, and 1.46$\times$, respectively, with respect to \NONESTED.
This means that \OPTIF is on average 1.95$\times$ faster than \MTBLAS over the 60 sparse matrices we consider.
The speed-up of \NESTED averages 0.90$\times$ over the 53 matrices for which it terminates.
All source files, raw data and scripts for re-doing this evaluation are available~\cite{github}.
%
%

%
%
%


%% file: related.tex

\vspace{0.2cm}
\textbf{Task-based Cholesky factorization.}
There is recent work on multi-threaded algorithms for sparse matrix factorization. 
In particular, Tang et al.~\cite{tangHybrid} developed a
multi-threaded version of CHOLMOD for hybrid architectures, using both CPU and GPU. 
Their speed-up is high thanks to the use of the GPU but 
their parallel implementation corresponds to the use of \outert tasks only. Davis et al.~\cite{davisGPU} also design a method for
sparse Cholesky factorizations on GPUs, using subtrees and batching to take advantage of the BLAS and LAPACK kernels on a GPU.
Hogg et al.~\cite{HoggCholesky} designed an algorithm for sparse Cholesky factorization using OpenMP.
They divide supernodes in blocks to have a really fine-grained implementation but their algorithm does not change
its granularity depending on the tasks and/or the input matrix as we do. Other known implementations include
MUMPS~\cite{mumps}, which is designed for distributed-memory machines with MPI, PARDISO~\cite{PARDISO}, which is a supernodal algorithm, and WSMP~\cite{WSMP}, which uses Pthreads.
Authors of \textsc{PaStiX}~\cite{pastixRuntime}
also investigated the use of a runtime-driven supernodal Cholesky factorization. 
Starting from the original scheduling scheme
of \textsc{PaStiX}~\cite{pastix}, which uses a 1D decomposition of the matrix similar to our \NONESTED strategy, they study the dynamical decomposition of the updates to
other supernodes into several nested tasks to accelerate the critical path of the algorithm, mainly for heterogeneous platforms. 
\textsc{PaStiX} does not take into account the sparsity of the matrix when creating nested tasks. 
It is thus equivalent to the \NESTED strategy. Note that in~\cite{pastix}, the authors use
timing models for BLAS kernels instead of considering the number of flops, but these timing
models need to be calibrated just as our algorithm \OPT and the algorithm was not task-based at the time.

\vspace{0.2cm}
\textbf{Multifrontal algorithms.}
Multifrontal methods constitute another class of algorithms targeting matrix factorization problems. 
These methods divide the input matrix into several blocks, called fronts, that are operated following the order defined by the elimination tree.
Contrarily to supernodal algorithms, the fronts are computed and moved along the elimination tree, instead of directly factorizing the supernodes.
As such, these methods exploit the parallelism available at the elimination tree level.
Multifrontal methods were introduced by Duff and Reid~\cite{duffreid} in the early 80s.
Other works introduced node amalgamation~\cite{duffreid,ashcraftSparse}, which consists in merging elimination tree nodes, creating bigger fronts and allowing more reuse of the data. It incurs additional storage and computation cost.
Some early multifrontal and parallel sparse Cholesky factorizations were proposed in the early 90s~\cite{schreiberParallel}, but
they were obviously not designed to run on modern architectures, even if node amalgamation was already in place.
A great survey about techniques used for general multifrontal algorithms was written in early 90s~\cite{liuMultifrontal}.
The work from Geist and Ng~\cite{GeistNgLeverage} describes an algorithm to leverage the cost of the subtrees in the elimination tree for Cholesky factorization.
The algorithm finds a separation layer in the tree to reach an acceptable load balance (in terms of flops). 

\vspace{0.2cm}
\textbf{About QR and LU factorizations.}
Recent parallel matrix factorizations using multifrontal algorithms exist for the QR and LU methods.
The method of Geist-Ng was re-used by
L'Excellent and Sid-Lakhdar~\cite{mumps-shared} for LU factorization, while also adding the use of multi-threaded BLAS calls for the top nodes of the tree. They also
switch from flops to estimated running times using models to avoid load imbalance due to the presence of many small fronts (where speed is limited) on one side and a few large fronts on another (where implementations
are more efficient). However, their approach corresponds to using only non-nested tasks or multi-threaded BLAS for one node, depending on its depth in the elimination tree, and we have shown that using multi-threaded BLAS
usually undermines the performance compared to a task-based approach in most cases.
Davis et al.~\cite{Davis2009MultifrontralQR} provide an
implementation for a rank-revealing QR using both parallelism in BLAS kernels and at tree level using Intel's Threading Building Blocks software.
Another fine-grained implementation of QR similar to our \NESTED strategy has been proposed~\cite{buttariQR}. It uses OpenMP.
This proposal assigns the creation and assembly of the fronts to independent tasks, while we assign the equivalent operations in the context of CHOLMOD to \outert tasks.
Previous approaches~\cite{mumpsQR} analyze the effect of the granularity on the performance for task-based QR factorization, based on 1D or 2D decomposition of the matrix.
However, they do not provide an
auto-adaptive algorithm: they run the algorithm with different parameters and report the best set for each input.
For LU, Booth et al.~\cite{BoothBasker} developed a parallel implementation based on the Gilbert-Peierls algorithm~\cite{GilbertPeierls}. Their approach
is not based on the use of supernodes and their implementation used Kokkos~\cite{kokkos} as a portable framework to provide parallelism. Closer to our work in term of implementation,
Ghysels et al.~\cite{HSS-Ghysels} use OpenMP to define a task-based implementation of multi-frontal LU factorization using different levels of nesting. They implement their own task-based linear algebra kernels so that
the node parallelism does not conflict with the tree parallelism. To limit the number of tasks created, they simply take into account the depth of a node in the elimination tree so that the deeper
nodes are merged with their parents.
Finally, Tang~\cite{tangPHD} also implemented the LU factorization for hybrid architectures.


%% file: conclusion.tex
%

This paper presents a selective nesting approach to drive the parallel granularity of CHOLMOD, a sparse Cholesky factorization algorithm, using OpenMP.
Our parallel implementation has two different types of tasks, \outert and \inner. The latter can be used to parallelize even more the former.
To deliver the best parallel approach, we present two heuristics that determine the \outert tasks for which it is better to use nested \inner tasks.
One of these heuristics, \OPT, determines a good nesting threshold for any input matrix, so that the user does not have to tune the algorithm beforehand.
It gives a very good performance for large matrices compared to the multi-threaded version of CHOLMOD (\MTBLAS), which relies on multi-threaded BLAS kernels.
It also proves to be better than the two extreme and more classical options of having either always or never nested tasks,
as the matrix sparse pattern has to be taken into account for creating the task-graph.
To further increase the generality of \OPT, we present an improved algorithm, \OPTIF. By setting
a threshold on the cost of the \inner tasks, we decrease the execution times by avoiding the creation of small tasks.
Both these heuristics run a fast algorithm before factorizing the input matrix to decide which parallel approach should be used.
Results 
delivered
by \OPTIF are similar to \OPT for large matrices and much better for small matrices. 
The average speed-up achieved by \OPTIF compared to \NONESTED
is 1.07$\times$ for small matrices and 1.83$\times$ for large matrices, while \MTBLAS delivers very low performance (down to 0.15$\times$ speed-up) for some matrices. Our whole evaluation was done using OmpSs and A64FX processors.

In the future, we plan to apply our algorithms to matrix factorizations running on x86 architectures.
Due to the high memory-bandwidth of A64FX processors, we expect the algorithm to be tuned to target a lower number of tasks.
In addition, we plan to further improve the performance by enabling multi-threaded BLAS and LAPACK kernels at the end of the factorization only, as tree parallelism is reduced when reaching the root.
